\documentclass[12pt]{article}%
\usepackage{amssymb}
\usepackage{amsfonts}
\usepackage{amsmath}
\usepackage[nohead]{geometry}
\usepackage[singlespacing]{setspace}
\usepackage[bottom]{footmisc}
\usepackage{indentfirst}
\usepackage{endnotes}
\usepackage{graphicx}%
\usepackage{rotating}
\makeatletter
\def\@biblabel#1{\hspace*{-\labelsep}}

\def\section{\@startsection {section}{1}{\z@}{-3.5ex plus -1ex minus 
    -.2ex}{2.3ex plus .2ex}{\large\bf}}

\def\subsection{\@startsection{subsection}{2}{\z@}{-3.25ex plus -1ex minus 
   -.2ex}{1.5ex plus .2ex}{\large\it}}
\makeatother
\begin{document}
\title{Automated Liquidity Provision and the Demise of Traditional Market Making\thanks{We thank Nadima El-Hassan, Bruce Lehmann, Danny Lo, Karyn Neuhauser, the participants of the \emph{Market Design and Structure} workshop at the Santa Fe Institute, and seminar participants at UTS for their helpful comments and suggestions.  This work was supported by the U.S. National Science Foundation Grant No. HSD-0624351 }} 
\author{\medskip\\
\medskip\\
\medskip\\Austin Gerig\thanks{Corresponding author.}\vspace{.05in}\\
School of Finance and Economics\\ University of Technology, Sydney\vspace{.05in}\\
austin.gerig@uts.edu.au\\
\medskip\\
\medskip\\
\medskip\\
David Michayluk\vspace{.05in}\\
School of Finance and Economics\\ University of Technology, Sydney\vspace{.05in}\\
david.michayluk@uts.edu.au\medskip\\
\medskip\\
\medskip\\}

\maketitle
\thispagestyle{empty}

\newpage
\thispagestyle{empty}

\doublespacing

\mbox{}\\
\begin{center}
\textbf{\large Abstract}
\end{center}
Traditional market makers are losing their importance as automated systems have largely assumed the role of liquidity provision in markets.  We update the model of Glosten and Milgrom (1985) to analyze this new world: we add multiple securities and introduce an automated market maker who uses the relationships between securities to price order flow.  This new automated participant transacts the majority of orders, sets prices that are more efficient, and increases informed and decreases uninformed traders' transaction costs.  These results can explain the recent dominance of high frequency trading in US markets and the corresponding increase in trading volume and decrease in transaction costs for US stocks.  
\mbox{ }


\pagebreak%
\setcounter{page}{1}

Traditionally, stock exchanges have appointed specialists or market makers to keep orderly markets and continually supply liquidity for specific securities.  For many years, these individuals played a key role in determining prices and participated in a large fraction of trading.  Today, this is no longer the case.  If you were to transact in the market right now, the counterparty to your trade is more likely to be a computer than a human.  Over the past decade, the task of liquidity provision has largely shifted from traditional market makers to proprietary automated systems that trade at high frequency and across different exchanges and securities.

Why are automated systems replacing traditional market makers, and should we be worried about this development?  Does this type of trading benefit the economy, or does it manipulate prices and take advantage of unsophisticated investors?

These questions are motivated by recent developments in US stock markets.  The specialists' fraction of trading volume at the NYSE has dropped precipitously in the last decade.\footnote{For reference, see Figs.~1,3 in Hendershott and Moulton (2009) where the specialists' fraction of NYSE volume declines from approximately $16\%$ to $2.5\%$ over the period January 1999 to May 2007.  Total floor trading itself went from approximately $52\%$ to $8\%$ over the same period.}  Contemporaneous with this decline has been a large and steady increase in high frequency (or low latency) automated trading.\footnote{The TABB Group reports that high frequency trading went from 21\% of market share in 2005 to 61\% in 2009.  See ``US Equity High Frequency Trading: Strategies, Sizing and Market Structure,'' September 2, 2009, available at http://www.tabbgroup.com.}  This type of trading is dominated by a handful of skilled firms who provide liquidity to the market using proprietary algorithms which, they claim, benefit investors by reducing transaction costs and making prices more efficient.  Given that these firms transact the majority of orders in the market, yet remain secretive and make large profits in undisclosed ways, it is not surprising that skeptics have questioned the integrity of their trading strategies and have called for government oversight of their activity.\footnote{See for example the recent article ``High frequency trading: Why the robots must die,'' May 7, 2010, available at http://wallstreet.blogs.fortune.cnn.com.}  

In this paper, we develop a model that can explain several characteristics of high frequency automated liquidity provision: why we should expect this type of trading to exist, why firms with highly skilled employees dominate the space, why they trade in large volumes, why prices are more efficient as a result of their trading, and finally who benefits from their presence.  With a simple modification of the model, we also can explain how automated liquidity provision leads to increased trading volume and decreased overall transaction costs.  Both of these trends have recently been observed for US stocks.\footnote{See Angel, Harris, and Spatt (2010) for an overview of the recently observed changes in US markets.}

The intuition behind the model is straightforward.  If traditional market making is described as connecting investors who transact in the same security at different times, then this new type of automated liquidity provision can be described as connecting investors who transact simultaneously at different exchanges and/or in different, but related, securities.  For example, if two securities XYZ and ZYX are similar to each other, then a buyer in XYZ and a seller in ZYX who are trading at the same time can be connected through the trades of a liquidity provider willing to take the opposite side of each order.  This liquidity provider can undercut the prices set by traditional market makers because the opposite direction of the trades makes it more likely one or both of the investors are uninformed.\footnote{If both traders where informed, why would one be buying and the other one selling?  Because XYZ and ZYX are similar, informed trading in these securities is more likely to be observed as contemporaneous buying or selling in both rather than buying in one and selling in the other.}  The opposite direction of the trades also allows the liquidity provider to offset his/her inventory risk (although we do not specifically analyze this effect here).

There are two main assumptions we make in the model.  First, we assume that automated liquidity providers set prices using information about all order flow in the market, but that traditional market makers only use information about order flow in the security they trade.  This captures the main advantage that machines have over their human counterparts: they can quickly and accurately process large amounts of relevant information when setting prices.  In real markets, we would not expect this assumption to strictly hold; human market makers are known to use trading activity in related securities to help determine their prices.\footnote{The results we present hold whether this assumption is strictly true or only approximately true, but are easier to present when strictly true.  Many of the results also hold if the superior information of automated liquidity providers was something else relevant to prices.  For example, automated systems could be better at accurately determining prices based on information about the state of the aggregate orderbook or the likelihood that an observed imbalance in buying and selling continues, etc.  We focus on cross-security information because the relevance of this information to prices is uncontroversial and easily modelled.}  Our point is simply that automated systems are superior at this task.

The second key assumption we make is that profitable investors can be separated into two types.\footnote{In the model, this separation is enforced by the dealership structure of the market.}  Investors of the first type are informed investors who are able to correctly value specific securities, possibly through superior analysis or because they are networked into information flows about these securities.  Investors of the second type are `quant' traders who are able to infer and correctly quantify the complex relationships between the values of different securities---they need not understand how or why these relationships exist.  If these different investor types exist and are separate, then information is incorporated into prices in the following way: it arrives through the trading activity of informed investors and is fully processed and applied to the rest of the market by `quant' traders.  These ideas are related to arbitrage pricing theory and statistical arbitrage.\footnote{See the papers by Ross (1976) and Gatev, Goetzmann, and Rouwenhorst (2006).  The correspondence between high frequency trading and statistical arbitrage is made by high frequency firms themselves.  See the commentary from Tradeworx, Inc. entitled ``Tradeworx, Inc. Public Commentary on SEC Market Structure Concept Release,'' April 21, 2010, available at http://www.tradeworx.com.}  To obtain clean results, we assume that the `quant' traders are strict liquidity providers who only transact against market orders, and that they are competitive such that their profits are driven to zero. In reality we would expect these traders to also trade aggressively (by using market orders) and to make sufficient profits for their existence.

The model we detail below is an extension of Glosten and Milgrom (1985).  Instead of using a single security, we include multiple securities, and we introduce a new type of liquidity provider called an \emph{automated market maker}.  This individual is our implementation of the `quant' trader discussed above.  We assume the automated market maker is competitive such that she requires no price concession and makes zero expected profit.  We show that in this case, she transacts the majority of order flow and that traditional market makers are largely priced out of the market.  This does not occur in an obvious way because average spreads remain unchanged.  Instead, it occurs because the automated market maker is better able to distinguish informed from uninformed order flow and therefore sets prices more precisely.  This means that prices are more efficient due to her activity, and that transaction costs are increased for informed investors and decreased for uninformed investors.  By extending the model to allow uninformed investors to trade more readily when their transaction costs are low, we also observe increased trading volumes and decreased overall transaction costs.

There are numerous studies that are related to this work.  Informed trading in a multi-asset setting has been modeled in several papers as an extension of the one-asset equilibrium model of Kyle (1985).\footnote{See Caball\'{e} and Krishnan (1994), Bernhardt and Taub (2008), and Pasquariello and Vega (2009).}  In these papers, the relationships between securities' prices are fully specified by the covariance matrix of returns, and all participants have knowledge of this matrix.  Here, we assume securities are related in a more general way, and we only allow the automated market maker to understand and/or act on these relationships.  This has two benefits.  First, it replicates the realistic scenario that sophisticated automated systems are needed to collect and instantaneously process order flow information from the entire market.  Second, it provides a natural explanation for the existence of statistical arbitrage opportunities in markets: that they result from the inability or disinterest of informed investors to apply their security-specific information to the rest of the market.  

There are other models that are similar in spirit to our work.  For example Boulatov, Hendershott, and Livdan (2010) and Chordia, Sarkar, and Subrahmanyam (2009) study how information is incorporated into a market with multiple related securities.  They develop models where informed trading in one security or type of security leads to cross-autocorrelations in the market.  Our model includes correlated price changes, but we do not observe cross-autocorrelations because we assume information is processed and applied to the market instantaneously by the automated market maker.  If this assumption was relaxed for some securities, then cross-autocorrelations would result.

The cost of liquidity provision has been described in various ways in previous papers and can be subdivided into three main categories: order-handling costs, inventory costs, and adverse selection costs (see Biais, Glosten, and Spatt (2005) for an overview).  In this paper, we focus only on the adverse selection component and do not consider order-handling or inventory costs.  Similar results would hold if these other costs were considered.  For example, Andrade, Chang, and Seasholes (2008) develop a multi-asset equilibrium model where risk-averse liquidity providers accommodate noninformational trading imbalances.  They show that demand shocks in one security will lead to cross-stock price pressure due to the hedging desires of liquidity providers, even without considering information effects.  

Our theoretical results nicely mirror recent empirical findings by Hendershott and Riordan (2009).  They report that computer generated trades contribute more to price discovery than human trading, and that computer generated quotes are at more efficient prices than human quotes.  They suggest that improvements in the speed and quality of access to electronic order books has spurred the use of automated trading.  We suggest that information about the relationships between securities can also improve computer generated price discovery.

The structure of the paper is as follows: Section I presents the model, Section II analyzes a two security market, Section III provides an example, Section IV extends the model, and Section V concludes.

\clearpage

\begin{center}
\section{The Model}
\end{center}

As in Glosten and Milgrom (1985), we assume a pure dealership market where \emph{informed investors} and \emph{liquidity traders} submit unit-sized buy and sell orders for a security with a random end-of-period value.  These orders are priced and cleared by a competitive, risk neutral \emph{market maker}.  Unlike the original Glosten and Milgrom model, we include multiple securities---each with a separate market maker and separate informed investors and liquidity traders---and introduce a new type of competitive, risk neutral liquidity provider called an \emph{automated market maker}.  The automated market maker is unique in that she is the only market participant who trades in multiple securities and understands how their end-of-period values are related; everyone else is unaware of this, unable to model it correctly, or does not have the sophistication to trade broadly and quickly on this information. 

Each round of trading takes place in two steps.  In the first step, one unit-sized order is submitted for each of the $N>1$ securities in the market.  These orders arrive randomly and anonymously from the pool of liquidity traders and informed investors within each security.  Liquidity traders are uninformed of the end-of-period value for the security they trade, are equally likely to buy or sell this security, and are always willing to accept the best price set by the liquidity providers.\footnote{This is the case of perfectly inelastic demand in the original Glosten and Milgrom (1985) model.  We relax this assumption in Section IV.}  Informed investors know with certainty the end-of-period value for the security that they trade, and they submit a buy or sell order based on this information.  If the end-of-period value is above the expected transaction price, they will place a buy order, and if it is below the expected transaction price, they will place a sell order.  The probability that an order for security $i$ is from an informed investor is public knowledge, is assumed to be larger than zero but less than one, and is denoted $\gamma_i$.  A buy order for security $i$ is denoted $B_i$ and a sell order is denoted $S_i$.  

In the second step, liquidity providers observe the set of orders placed, determine the prices at which they are willing to transact, and then transact with an order if their price is the most competitive.  If the market maker for security $i$ and the automated market maker set the same price for a transaction, then we assume each partakes in half of the trade.  Notice that we assume prices are set by liquidity providers after all orders have been placed, whereas in the original Glosten and Milgrom (1985) model, quotes are set by the liquidity provider before transactions take place.  In real markets, transactions occur randomly in time across securities, and quotes for any particular security will be dynamically updated as transactions occur in other related securities.  For simplicity, we assume these updates are collapsed into one moment in time.

We assume that the true value of security $i$ can take on one of two possible values at the end of the trading period, $\tilde{V}_i=V_i$, where $V_i\in\{V_i^+,V_i^-\}$ and,
\begin{eqnarray}
V_i^+ & = & p_i + r_i,\\
V_i^- & = & p_i - r_i.
\end{eqnarray}
$r_i$ is a constant that sets the scale of price changes for security $i$.  For simplicity, we assume that each of these occur with probability 1/2 so that the unconditional expected value of security $i$ is $p_i$.  Everyone in the model is aware of these possible outcomes and their probabilities.

We assume that the value of securities are related to one another such that,
\begin{equation}
P(V_i) \neq P(V_i|V_j)\neq P(V_i|V_j,V_k) \neq \dots  \neq P(V_i|V_j,V_k, \dots, V_N).
\end{equation}
This information is not known to the traders or the market makers, but it is known to the automated market maker and used to price transactions.

In general, the end of the trading period can occur at some distant point in the future, so that there can be multiple rounds of trading before the end-of-period value for each security is revealed.  To simplify the analysis, and because our main results hold for only one round of trading, we only analyze the case where the trading period ends after one round.  When presenting results, we implicitly assume that many of these trading periods have taken place and we take averages over the outcomes.

Because liquidity providers are competitive and risk neutral, they set prices such that they expect to make zero profit on their trades (see Glosten and Milgrom (1985)).  For example, in a market with only one security, both the market maker and the automated market maker would set the price of an order at the expected value of the security given the order,
\begin{eqnarray}
E\left[\tilde{V}_1\middle|B_1\right] & = & p_1 + \gamma_1 r_1, \\
E\left[\tilde{V}_1\middle|S_1\right] & = & p_1 - \gamma_1 r_1.
\end{eqnarray}
Therefore, a buy order transacts at price $p_1 + \gamma_1 r_1$ and a sell order at price $p_1 - \gamma_1 r_1$ in this scenario.  Because orders are cleared at these prices, there is no expected profit for the liquidity providers.  

In the propositions below, we assume that everyone is aware of the setup of the model and has traded long enough to know what actions are optimal for them.  After stating the propositions and giving a short explanation of each, we prove them for a two security market.  Proofs for markets with more than two securities are straightforward extensions of these proofs and are available on request.\newline

\noindent {\bf PROPOSITION 1.} {\it The market maker for security $i$ transacts a minority of order flow and competes with the automated market maker only on the widest spreads.}\newline

The market maker cannot compete with the automated market maker on pricing because he is not using information about the relationships between securities.  If he sets prices to their expected value given his information, he will find himself transacting unprofitable orders and not transacting profitable ones.  This is because the automated market maker knows which orders are profitable at these prices; she takes the profitable ones away from the market maker by pricing them more aggressively and leaves the unprofitable ones by pricing them less aggressively.  Because the market maker is transacting unprofitable orders (which contain a larger fraction of informed trades), he must widen his spread in price between buy orders and sell orders.  The equilibrium point for him is not reached until he is using the overall highest and lowest prices set by the automated market maker.\newline

\noindent {\bf PROPOSITION 2.} {\it With the addition of the automated market maker, the spread in average transaction price between buys and sells in security $i$ remains the same, but is smaller for liquidity traders and larger for informed investors.}\newline

The automated market maker is unable to reduce spreads because the unconditional probability of an order arriving from an informed versus a liquidity trader for each security remains unchanged.  She, however, can use information about the orders placed in other securities to help distinguish if a particular order is likely from an informed investor vs. a liquidity trader.  She uses this information to price these situations differently.  Informed investors receive worse prices because they now compete with one another across securities; this means the average spread in price between a buy and a sell increases for them.  Liquidity traders receive better prices because their trades can be connected to other liquidity traders in different securities through the actions of the automated market maker.  For these traders there is a decrease in the average spread in price between buys and sells.\newline

\noindent {\bf PROPOSITION 3.} {\it With the addition of the automated market maker, prices are more efficient, i.e., on average, the transaction price for security $i$ is closer to its end-of-period value.}\newline

The automated market maker uses a better information set than that used by the market maker and can therefore price order flow more precisely.  This means the transaction price is, on average, closer to the security's end-of-period value.  This effect is most pronounced in securities that have a small proportion of informed trading.

In the original Glosten and Milgrom (1985) paper, as the liquidity provider observes more order flow, transaction prices approach the security's fundamental value.  This proposition documents the same effect, but now it occurs because the automated market maker can observe `order flow substitutes' for security $i$ by observing the orders for the rest of the securities in the market.

{\centering
\section{Two Security Market}
}

In what follows, we prove the propositions for a two security market where $P(V_1^+|V_2^+)=\phi$.

\subsection{Proposition 1}

\begin{figure}
\begin{center}
\includegraphics[width=5in]{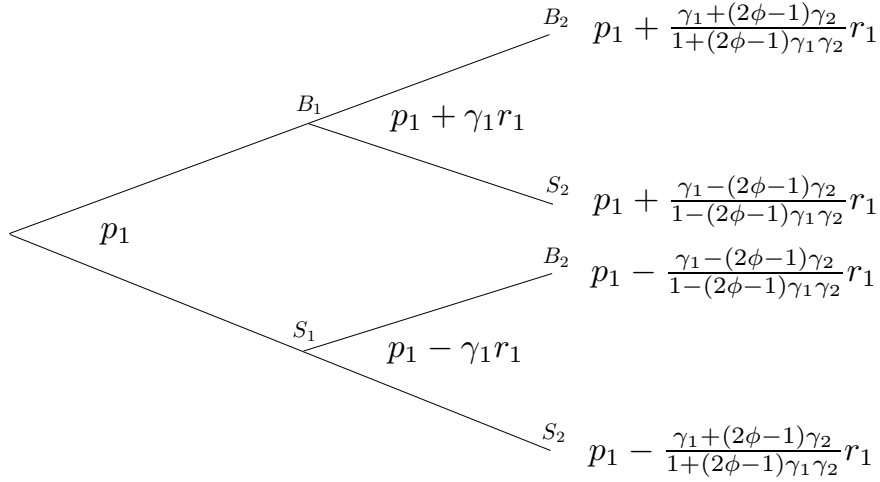}
\end{center}
\caption{\label{fig.tree}Diagram of the transaction prices for security 1 in the 4 possible order flow states, $(B_1,B_2), (B_1,S_2), (S_1,B_2), (S_1,S_2),$ in a two security market.  The values shown at earlier nodes are not transaction prices, but are the expected price at those nodes.}
\end{figure}

There are four possible order flow states in this market, $(B_1,B_2), (B_1,S_2), (S_1,B_2), (S_1,S_2)$.  Because the values of the two securities are related to each other, and because informed investors trade in both, certain order flow states will be observed more often than others.  For example, when the values of security 1 and security 2 are positively correlated, then the states $(B_1,B_2)$ and $(S_1,S_2)$ are more likely to occur, and the orders in these states are more likely to be informed.  The automated market maker is aware of this and sets prices accordingly.  She sets the transaction price of orders for security 1, $T_1$, to the expected value of security 1 conditioned on the particular order flow state.  This is calculated as follows,
\begin{eqnarray}
T_1(B_1,B_2)=E\left[\tilde{V}_1\middle|B_1,B_2\right]& = & p_1+\frac{\gamma_1+(2\phi-1)\gamma_2}{1+(2\phi-1)\gamma_1\gamma_2}r_1,\label{eq.price1} \\
T_1(B_1,S_2)=E\left[\tilde{V}_1\middle|B_1,S_2\right]& = & p_1+\frac{\gamma_1-(2\phi-1)\gamma_2}{1-(2\phi-1)\gamma_1\gamma_2}r_1,\label{eq.price2} \\
T_1(S_1,B_2)=E\left[\tilde{V}_1\middle|S_1,B_2\right]& = & p_1-\frac{\gamma_1-(2\phi-1)\gamma_2}{1-(2\phi-1)\gamma_1\gamma_2}r_1,\label{eq.price3} \\
T_1(S_1,S_2)=E\left[\tilde{V}_1\middle|S_1,S_2\right]& = & p_1-\frac{\gamma_1+(2\phi-1)\gamma_2}{1+(2\phi-1)\gamma_1\gamma_2}r_1.\label{eq.price4}
\end{eqnarray}
The transaction prices in Eqs.~\ref{eq.price1}-\ref{eq.price4} are shown in diagram form in Fig.\ref{fig.tree}.  The market maker for security 1 is unaware of the importance of conditioning prices on the order in security 2 (or is simply unable to do so), and therefore does not use Eqs.~\ref{eq.price1}-\ref{eq.price4} to price order flow.  He \emph{is} aware of the following,
\begin{eqnarray}
E\left[\tilde{V}_1\middle|B_1\right] & = & p_1 + \gamma_1 r_1, \label{eq.price5}\\
E\left[\tilde{V}_1\middle|S_1\right] & = & p_1 - \gamma_1 r_1. \label{eq.price6}
\end{eqnarray}
The market maker would set prices at these values if the automated market maker were not present.

In what follows, we assume that $\phi$ is restricted to $(1/2,1]$, such that the values of security 1 and 2 are positively correlated.  The results would be similar, although with some signs and descriptions reversed, if they were negatively correlated.  The propositions hold in either case.  Because $\gamma_1$ and $\gamma_2$ are restricted to $(0,1)$, it is straightforward to show that,
\begin{equation}
E\left[\tilde{V}_1\middle|B_1,B_2\right] > E\left[\tilde{V}_1\middle|B_1\right] > E\left[\tilde{V}_1\middle|B_1,S_2\right]
 \,^\leftarrow_\rightarrow\, E\left[\tilde{V}_1\middle|S_1,B_2\right] > E\left[\tilde{V}_1\middle|S_1\right] > E\left[\tilde{V}_1\middle|S_1,S_2\right],\label{eq.equality}
\end{equation}
where $^\leftarrow_\rightarrow$ denotes that $E\left[\tilde{V}_1\middle|B_1,S_2\right]$ and $E\left[\tilde{V}_1\middle|S_1,B_2\right]$ may be switched in this ordering.  Notice that the automated market maker sets the price for $B_1$ more aggressively when it is accompanied by $S_2$ and more timidly when accompanied by $B_2$ (with the opposite results for $S_1$).  The automated market maker realizes that the opposite direction of orders makes it more likely both investors are uninformed and that the same direction makes it more likely they are informed, and she prices them accordingly.  Because the market maker does not realize this, his role is marginalized.  Suppose the market maker tried to set prices according to Eqs.~\ref{eq.price5}-\ref{eq.price6}.  If he transacted a random sampling of orders at these prices, he would make zero expected profit.  Unfortunately for him, the automated market maker ensures that his transactions are not a random sampling.  She transacts orders that would have been profitable to him by pricing them more aggressively (this occurs for $(B_1,S_2)$ and $(S_1,B_2)$, see Eq.~\ref{eq.equality}), and she allows him to transact orders that are unprofitable because she prices them less aggressively (this occurs for $(B_1,B_2)$ and $(S_1,S_2)$, see Eq.~\ref{eq.equality}).  Because the market maker is now receiving a larger fraction of informed trading, he is forced to widen his spread until he expects to make zero profit.  This occurs at the widest spread set by the automated market maker. This same argument applies to the market maker for security 2.   

We have assumed that when the market maker and automated market maker set the same price for a transaction, they each partake in half of the trade.  This means that the fraction of order flow transacted by the market maker is equal to 1/2 of the proportion of order flow at the widest spread.  This is,
\begin{equation}
P(B_1,B_2)/2+P(S_1,S_2)/2 = \left[1+(2\phi-1)\gamma_1\gamma_2\right]/4.
\end{equation}
Because $\gamma_1$ and $\gamma_2$ are restricted to $(0,1)$ and $\phi$ is restricted to $(1/2,1]$, the maximum value of this equation is less than 1/2.  Therefore, the market maker in security 1 trades a minority of order flow.  The same argument applies to the market maker for security 2.

\subsection{Proposition 2}

\begin{figure}
\begin{center}
\includegraphics[width=6in]{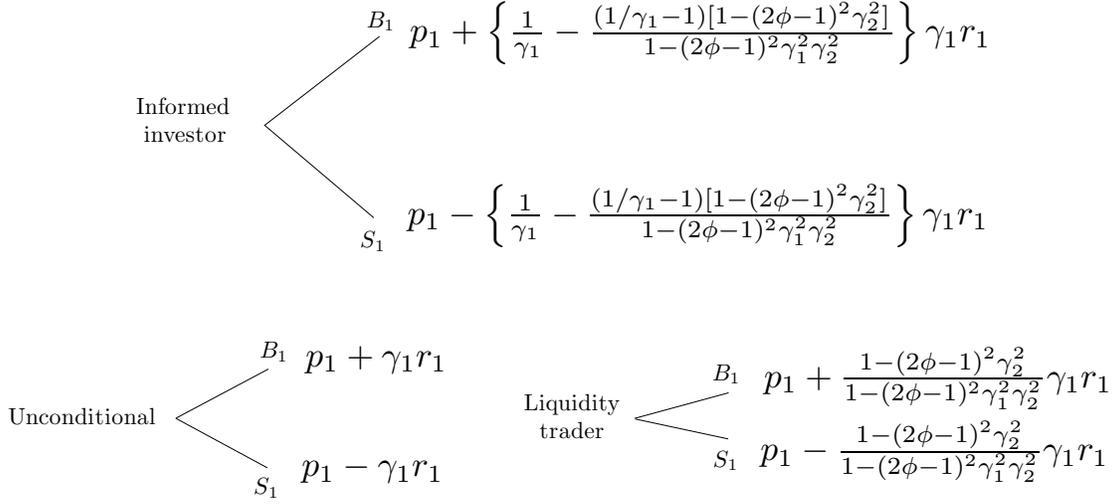}
\end{center}
\caption{\label{fig.spread}Diagram of the average buy and sell transaction price for informed investors, liquidity traders, and unconditional on trader type.  Notice that informed investors realize a larger spread than liquidity traders, and that the spread without the automated market maker is the expected spread over both of these trader types.}
\end{figure}

We can calculate the expected transaction price of informed investors when buying and selling security 1 as follows,
\begin{eqnarray}
E\left[T_1|B_1, I_1\right] & = & E\left[T_1\middle|B_1, B_2, I_1\right] P(B_2|B_1, I_1) + E\left[T_1\middle|B_1, S_2, I_1\right] P(S_2|B_1, I_1), \\
E\left[T_1\middle|S_1, I_1\right] & = & E\left[T_1\middle|S_1, S_2, I_1\right] P(S_2|S_1, I_1) + E\left[T_1\middle|S_1, B_2, I_1\right] P(B_2|S_1, I_1),
\end{eqnarray}
where the conditioning variable $I_1$ means the order for security 1 was from an informed investor.  The expected spread in price between buying and selling security 1 for an informed investor, $\Delta_{1,I}$, is therefore,
\begin{equation}
\Delta_{1,I} = 2 \gamma_1 r_1\left\{\frac{1}{\gamma_1} - \frac{(1/\gamma_1-1)[1-(2\phi-1)^2\gamma_2^2]}{1-(2\phi-1)^2\gamma_1^2\gamma_2^2} \right\} 
\end{equation}
In a similar way, we calculate the spread between buying and selling security 1 for a liquidity trader, $\Delta_{1,L}$,
\begin{equation}
\Delta_{1,L} = 2\gamma_1 r_1\left[\frac{1-(2\phi-1)^2\gamma_2^2}{1-(2\phi-1)^2\gamma_1^2\gamma_2^2} \right] 
\end{equation}
$\Delta_{1,I}$ and $\Delta_{1,L}$ are to be compared to the unconditional spread, $\Delta_1$, which can be calculated as follows,
\begin{eqnarray}
\Delta_1 & = & \gamma_1 \Delta_{1,I}  + (1-\gamma_1) \Delta_{1,L} , \\
 & = & 2\gamma_1 r_1.
\end{eqnarray}
This is just the spread that would be observed without the automated market maker (subtract Eq.~\ref{eq.price6} from Eq.~\ref{eq.price5}).  The same argument applies for security 2.  In Fig.~\ref{fig.spread}, we diagram these results.

Because $\gamma_1$ and $\gamma_2$ are restricted to $(0,1)$, and $\phi$ is restricted to $[0,1/2)$ or $(1/2,1]$, then it is straightforward to show that,
\begin{equation}
\Delta_{1,I} > \Delta_1 > \Delta_{1,L}
\end{equation}
so that the spread is increased for informed investors and decreased for liquidity traders. The same argument applies for security 2.

\subsection{Proposition 3}

To determine the inefficiency of transaction prices, we calculate the expected absolute difference between $\tilde{V}_1$ and the transaction price, $T_1$,
\begin{equation}
E\left[|\tilde{V}_1-T_1|\right]=E\left[|\tilde{V}_1-T_1|\middle|I_1\right]P(I_1) + E\left[|\tilde{V}_1-T_1|\middle|L_1\right]P(L_1).
\end{equation}
where $I_1$ and $L_1$ denote that the order in security 1 was placed by an informed investor and a liquidity trader respectively.  Calculating this, we have,
\begin{equation}
E\left[|\tilde{V}_1-T_1|\right] = (1-\gamma_1)\left[1+\frac{1-(2\phi-1)^2\gamma_2^2}{1-(2\phi-1)^2\gamma_1^2\gamma_2^2}\gamma_1\right]r_1.
\end{equation}
Without the automated market maker, the expected absolute difference between the end-of-period value and the transaction price is,
\begin{equation}
E\left[|\tilde{V}_1-T_1|\middle| \mbox{No Auto MM}\right] = (1-\gamma_1)(1+\gamma_1)r_1.
\end{equation}
Because $\gamma_1$ and $\gamma_2$ are restricted to $(0,1)$, and $\phi$ is restricted to $[0,1/2)$ or $(1/2,1]$, then it is straightforward to show,
\begin{equation}
E\left[|\tilde{V}_1-T_1|\right] < E\left[|\tilde{V}_1-T_1|\middle| \mbox{No Auto MM}\right],
\end{equation}
so that transaction prices in security 1 are closer to the end-of-period value for security 1 when the automated market maker is included.  The same argument applies for security 2.

{\centering
\section{An Example}
}

\begin{figure}
\begin{center}
\includegraphics[width=5in]{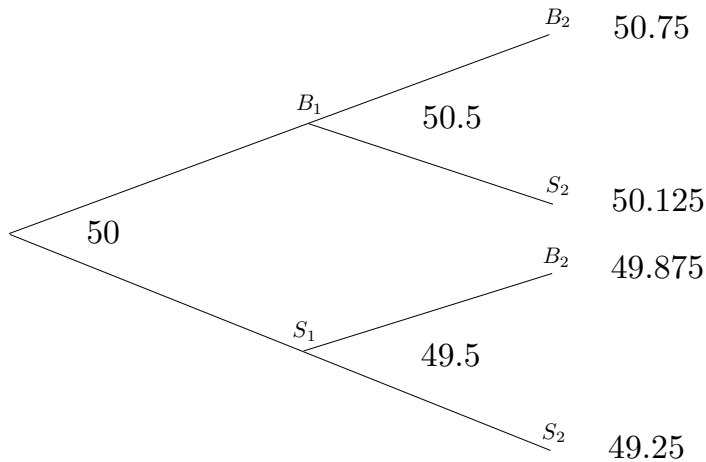}
\end{center}
\caption{\label{fig.tree2}Diagram of the transaction prices for security 1 in the 4 possible order flow states, $(B_1,B_2), (B_1,S_2), (S_1,B_2), (S_1,S_2),$ in a two security market.  The values shown at earlier nodes are not transaction prices, but are the expected price at those nodes.  The parameters used are: $\gamma_1=\gamma_2=.5$, $\phi=0.9$, $p_1=50$, and $r_1=1$.}
\end{figure}

\begin{figure}
\begin{center}
\includegraphics[width=6in]{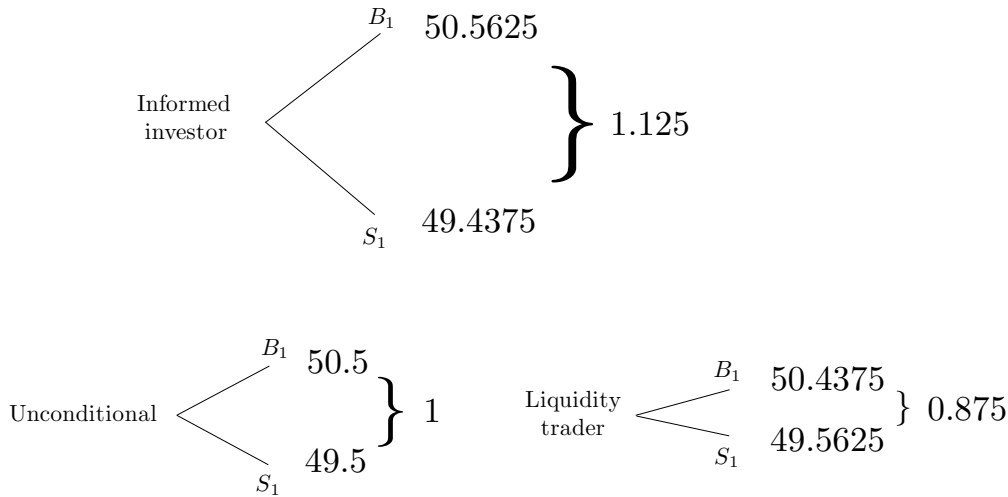}
\end{center}
\caption{\label{fig.spread2}Diagram of the average buy and sell transaction price for informed investors, liquidity traders, and unconditional on trader type.  Notice that informed investors realize a larger spread than uninformed investors.  The parameters used are $\gamma_1=\gamma_2=.5$, $\phi=0.9$, $p_1=50$, and $r_1=1$.}
\end{figure}

As an example, consider a two security market with the following parameters: $\gamma_1=\gamma_2=.5$, $P(V_1^+|V_2^+)=\phi=0.9$, $p_1=50$, and $r_1=1$.  Fig.~\ref{fig.tree2} shows the transaction prices set for security 1 by the automated market maker conditioned on different order flow states.  The market maker will price orders using only the highest and lowest price in the figure, which is $50.75$ for a buy and $49.25$ for a sell.  The proportion of order flow he transacts is equal to 1/2 the proportion of orders at the widest spread.  Using Eq.~13, this is 30\% of the orders for security 1.  This means the automated market maker transacts the other 70\%, and that the automated market maker transacts the majority of order flow.

The average spread in the price between buy orders and sell orders is 1 overall, 0.875 for liquidity traders, and 1.125 for informed investors.  These values are shown in Fig.~\ref{fig.spread2}.  Notice that the spread increases for informed investors and decreases for liquidity traders with the addition of the automated market maker.

The average difference between the transaction price and the end-of-period value of security 1 is 0.71875 when the automated market maker is present, and is 0.75 when she is not.  This means prices are more efficient with the addition of the automated market maker.

{\centering
\section{Extension of Model}
}

Whereas before we assumed that liquidity traders were always willing to buy or sell, we now extend the model to allow liquidity traders to refrain from trading if their expected transaction cost is too high.  Specifically we assume that the fraction of liquidity traders who are willing to submit an order is a monotonically decreasing function of their expected transaction cost.  This means the following two propositions hold:\newline

\noindent {\bf PROPOSITION 4.} {\it In the extended model, with the addition of the automated market maker, the probability of a transaction in security $i$ increases, i.e., expected volumes increase.}\newline

When the automated market maker is added to the market, the expected transaction cost of a liquidity trader is reduced (Proposition 2).  In the extension of the model, we have assumed this reduced cost increases the probability that a liquidity trader will place an order.  Trivially, this increases the overall probability of a transaction.\newline

\noindent {\bf PROPOSITION 5.} {\it In the extended model, with the addition of the automated market maker, the spread in average transaction price between buys and sells in security $i$ is reduced.}\newline

From the previous explanation for Proposition 4, we know that the addition of the automated liquidity provider increases the probability of a liquidity trader placing an order.  This decreases the fraction of orders that come from informed investors.  When the fraction of orders from informed investors decreases, adverse selection is reduced and liquidity providers can set smaller spreads for buy and sell orders.\newline

Below, we prove the new propositions for a two security market.  Proofs for markets with more than two securities are straightforward extensions of the two security case and are available on request.

\subsection{Proposition 4}

We denote the fraction of informed investors in security $1$ by $\delta_1$ and the fraction of liquidity traders who are willing to submit an order by $\pi_1$.  The probability of a transaction in security 1, $P_1$, is therefore,
\begin{equation}
P_1 = \delta_1 + (1-\delta_1)\pi_1.
\end{equation} 
The expected transaction cost of a liquidity trader is $1/2$ the expected spread, which is $\Delta_{1,L}/2$ (Eq.~17).  Without the automated market maker, this would be $\Delta_1/2$ (Eq.~18).  Because $\pi_1$ is a monotonically decreasing function of the liquidity traders' expected transaction costs, and because $\Delta_{1,L}/2<\Delta_1/2$, then $\pi_1$ is larger when the automated market maker is added.  Therefore, the probability of a transaction in security 1, $P_1$, is also larger when the automated market maker is added.  The same argument applies for security 2.

\subsection{Proposition 5}

The fraction of orders for security $i$ that come from informed investors is,
\begin{equation}
\gamma_1 = 1-(1-\delta_1)\pi_1. 
\end{equation}
As shown in the proof of Proposition 4, $\pi_1$ is larger when the automated market maker is added.  Therefore, from Eq.~26, $\gamma_1$ is smaller.  The unconditional average spread in price between buying and selling in security 1 is given in Eq.~19, $\Delta_1 = 2\gamma_1r_1$, and is therefore reduced when the automated market maker is added.  The same argument applies for security 2.

{\centering\section{Conclusion}}

The purpose of this paper is to explain the increasing dominance of automated liquidity provision and to understand its effects on the market.  In the model posited above, we show that automated liquidity providers are able to price securities more precisely than traditional market makers.  As a result, they transact the majority of order flow and cause prices to be more efficient.  This occurs not only because they trade across securities instantaneously, but also because their actions are determined by highly skilled individuals who accurately model the complex relationships between securities.  Traditional market makers simply cannot compete on either front.  

The presence of automated liquidity providers has material effects on investors in the market.  Informed investors make smaller profits and uninformed investors lose less money.  Informed investors make smaller profits because they must now compete with one another across securities.  Uninformed investors lose less money because they are able to transact through the liquidity provider to other uninformed investors in related securities.  If the uninformed increase their trading activity due to lower transaction costs, overall volumes increase and overall transaction costs are reduced.  These results match nicely with recently observed changes in US markets, where high frequency automated trading now dominates.

\singlespacing

\end{document}